\documentstyle[emulateapj,epsfig,apjfonts]{article}
\def\gs{\mathrel{\raise0.35ex\hbox{$\scriptstyle >$}\kern-0.6em 
\lower0.40ex\hbox{{$\scriptstyle \sim$}}}}
\def\ls{\mathrel{\raise0.35ex\hbox{$\scriptstyle <$}\kern-0.6em 
\lower0.40ex\hbox{{$\scriptstyle \sim$}}}}

\newcommand{\lya}{{Ly$\,\alpha$}}

\newcommand{\mum}{$\,\mu$m}
\newcommand{\pho}{$\phantom{1}$}
\newcommand{\Pho}{$\phantom{{-}}$}
\lefthead{Chapman et al.}
\righthead{Sub-mm imaging of a proto-cluster region at $z$=3.09}

\begin{document}

\title{Sub-mm imaging of a proto-cluster region at \lowercase{$z=3.09$}
}

\author{S.\,C.\ Chapman,$\!$\altaffilmark{1} G.\,F.\ Lewis,$\!$\altaffilmark{2}
D.\ Scott,$\!$\altaffilmark{3} E.\ Richards,$\!$\altaffilmark{4}
C.\ Borys,$\!$\altaffilmark{3} 
C.\,C.~Steidel,$\!$\altaffilmark{5} K.\,L.~Adelberger,$\!$\altaffilmark{5}
A.\,E.~Shapley$\!$\altaffilmark{5}
}
\affil{$^1$Observatories of the Carnegie Institution of Washington, Pasadena,
Ca 91101,~~U.S.A.}
\affil{$^2$Anglo-Australian Observatory, P.O. Box 296, Epping,
NSW 1710,~~Australia}
\affil{$^3$Department of Physics \& Astronomy,
        University of British Columbia,
        Vancouver, B.C.~V6T 1Z1,~~Canada}
\affil{$^4$Department of Physics \& Astronomy, Arizona State University,
P.O. Box 8701504, Tempe, Az 85287-1504,~~U.S.A.}
\affil{$^5$Palomar Observatory, Caltech 105-24,
        Pasadena, Ca 91125,~~U.S.A.}



\begin{abstract}
We have used the Submillimetre Common-User Bolometer Array (SCUBA) detector
on the  James Clerk Maxwell Telescope (JCMT)
to measure bright sub-mm emission associated with
a recently discovered extensive (${>}\,100h^{-1}$kpc) and highly luminous,
`blob' of Ly$\,\alpha$ emission at $z=3.09$. 
The blob lies within a known large overdensity
of optical sources in the $z=3.07$--3.11 range, and is centered on a
locally overdense peak within this region.
The best explanation for the copious sub-mm emission is a dust obscured
continuum source, which may produce the ionizing flux for the Ly$\,\alpha$
cloud. Cooling gas explanations are plausible but excessively complicated,
and the 450/850$\,\mu$m ratio rules out a significant 
fraction of the signal arising from the Sunyaev-Zel'dovich increment.  
At least two additional ${\simeq}\,10\,$mJy
sub-mm detections in the SCUBA map, with a surface density significantly higher
than in blank field surveys, suggests that they may be associated with
the $z\,{=}\,3.09$ structure. 
A SCUBA `photometry' observation of a second nearby Ly$\,\alpha$ blob
tentatively detects a weaker sub-mm counterpart. 
\end{abstract}


%
%
%
\section{Introduction}

Deep sub-mm surveys with SCUBA have detected distant luminous galaxies which
emit most of their energy at far infrared wavelengths
(e.g., Smail et al.~1997; Barger et al.~1998, Hughes et al.~1998,
Eales et al.~1998).
Although some of these sources have been identified with
very faint and sometimes very red optical counterparts (Smail et al.~1999)
their poorly constrained redshift distribution (Barger et al.~1999, 
Smail et al.~2000) has made their interpretation unclear.
Bright distant SCUBA sources appear similar to local ultra-luminous infra-red
galaxies (e.g.~Sanders 2000), but represent a more significant fraction
of the global star formation at high redshift.  How these differ at
high $z$, and how they are related to optically selected galaxies remains
to be understood.

We might expect such luminous sub-mm sources 
to be especially prevalent in regions that will one day
become rich clusters of galaxies.  For instance, if early type galaxies
(which are dominant only in rich cluster environments) form by 
major merger events, then during the epoch of their formation
they might produce a large fraction of the
bright sub-mm sources in the Universe.  There might therefore be
a strong clustering signal from sub-mm sources
at high redshift (e.g.~Ivison et al.~2000). 
Even if the sub-mm sources are primarily 
AGN-driven, we might expect them to be strongly clustered,
given the bulge to black hole mass correlations
(cf., Gebhardt et al.~2000).

Since it is not yet possible to measure the
clustering properties of sub-mm sources, the next best approach is
to examine regions that are already known to be overdense from other
techniques.  The most
likely time in a cluster's history for a bright sub-mm phase is when
smaller groups are collapsing, prior to virialization of the whole
proto-cluster environment, when compact groups of objects
with relatively low velocity dispersion present high probabilities
for merger events. Therefore, galaxy concentrations within known large-scale
overdensities are good places to search for significant sub-mm emission. 

Extensive spectroscopic surveys have 
recently shown strong clustering among the star forming Lyman break
galaxy (LBG) population at $z\,{\sim}\,3$ (e.g., Adelberger et al.~1998).   
In one striking example, 
a high-contrast redshift overdensity of LBGs at $z\,{=}\,3.09$ in the SSA22 
survey field was discovered (Steidel et al.~1998), with  
deep, narrow band \lya\ imaging used to identify at least 160 likely 
members of this structure (Steidel et al.~2000).
Most dramatic in this $z\,{=}\,3.09$ spike
was the discovery of two extensive (${>}\,100h^{-1}$kpc), highly luminous
($L_{{\rm Ly}\alpha}\,{\sim}\,10^{37}$W) `blobs' of Ly$\,\alpha$
emission. The `blobs' resemble the giant Ly$\,\alpha$ 
nebulae associated with high redshift radio galaxies 
(cf.\ van Ojik et al.~1996), but have ${<}\,$1\% of the associated 
radio continuum flux and no obvious source of UV photons bright 
enough to excite the nebular emission.  Fabian et al.~(1986)
suggested that such extended emission line nebulae might
be the signature of `maximal cooling flows' 
in  proto-cluster  and proto-galaxy environments -- these
`blobs' resemble just such objects.

The SSA22 $9\times9$ arcmin field at $z\,{=}\,3.07$--3.11 is
a factor of 6 overdense in LBGs relative to the general field, and about
twice as overdense again within the local density peak centered on `blob~1'.
At present, the SSA22 structure is by far the best studied candidate for
a clustered region at $z\,{>}\,1$.
The space density of the LBG redshift `spikes'
is similar to that of rich x-ray clusters
at low redshift, supporting the plausible identification of the LBG 
overdensities as proto-cluster sites.

We have obtained SCUBA 850$\mu$m and 450$\mu$m maps of the Ly$\,\alpha$ blob 
lying at an overdensity peak within the already overdense field,
revealing an extremely luminous sub-mm counterpart.
In this letter, we describe the available multi-wavelength
data for the blob regions and explore the relevant excitation mechanisms.

\section{Observations and Reduction}


The field centered on the Ly$\,\alpha$ `blob~1' 
(RA$_{2000}$ = $22^{\rm h}$ $17^{\rm m}$ $26^{\rm s}$,
$\delta_{2000}$ = $+00^{\circ}$ $12^{\prime}$ $40^{\prime\prime}$) 
was observed with the  SCUBA instrument (Holland et al.~1999)
on the JCMT.  During an observing run in May 2000, we
obtained 450 and 850\mum\ images
simultaneously in `jiggle' mode,
with half-power beam widths of 7.5 and 14.7 arcsec respectively.
The effective integration time (total time spent on the map, on and off 
source) was 8.2\,ks, reaching a 1$\sigma$ sensitivity 
of $3.3\,$mJy and $24\,$mJy at 850\mum\ and 450\mum\ respectively. 
We used 45 arcsec chop throws at fixed RA and declination
with position angles of 0, 90 and 135 degrees from north, to avoid consistently
chopping any structures onto each other. 
For the central source, the 850\mum\ flux density was measured by fitting a
single beam template (derived from observing Mars)
to the source in each scan and combining the results with 
inverse variance weighting.
The fluxes are consistent with each other in each of the individual scans.
Since no other sources are detected in the individual scans, 
we simply stacked and combined the scans 
and similarly fit beam templates to all sources representing 
peaks $>$2.8$\sigma$ in the final map.  
In addition, a `photometry' observation in three-bolometer chopping
mode (see Chapman et al.~2000a, Scott et al.~2000)
was performed on the \lya\ `blob~2'
for an effective on-source observing time of 2.8\,ks, 
reaching 
an RMS of 1.2\,mJy at 850\mum.

Pointing was checked hourly on blazars and a sky-dip was
performed to measure the atmospheric
opacity. The pointing variations were around 2 arcsec, while the
average atmospheric zenith opacities at 450\mum\ and 850\mum\ were
1.39 and 0.22.
The data were reduced using the  Starlink package SURF (Jenness et al.~1998).
Spikes were first carefully rejected from the
data, followed by correction for atmospheric opacity
and sky subtraction using the median of the outer ring array elements.
We then weighted each pixel by its timestream inverse variance,
relative to the central pixel.
The multiple scans taken at dithered positions, together with sky 
rotation throughout the observations, 
ensure that each point of the map is covered by several bolometers.
The data were finally calibrated against the
compact \hbox{H\,{\sc ii}~} region $16293{-}2422$, observed  during the
same nights. 

Archive VLA data were retrieved for this field, with
RMS sensitivities of 60$\,\mu$Jy at 1.4\,GHz and 40$\,\mu$Jy at 8.5\,GHz.
The beam is comparable to the SCUBA beam at 850\mum, but since the field
was observed as a low declination snapshot, the beam is
elongated to approximately $14\times20$ arcsec.

The visible images on which our present analysis is based
were obtained using
the Palomar 200\,inch telescope with the COSMIC camera.
Spectroscopy and near-IR imaging  were obtained with the Keck 10m
telescope and the LRIS and NIRC instruments respectively, and are
described elsewhere (Steidel et al.~2000).

\section{analysis and results} 

\subsection{The blob sources}
We have detected an unresolved  sub-mm counterpart to the 
`blob~1' (SMMJ\,221726+0013)
with an 850\mum\ flux density of 20\,mJy.  At $z\,{=}\,3.09$
this corresponds to a luminosity
$\nu L_\nu\,{=}\,3.4\times10^{11}h^{-2}{\rm L}_\odot$
for an $\Omega\,{=}\,1$ cosmology, giving a bolometric luminosity of
${>}\,10^{13}{\rm L}_\odot$ for any reasonable temperature and emissivity
index.  There is also a ${\sim}\,3\sigma$ detection at the central position
in the 450\mum\ map (76\,mJy), which also does not appear to be extended.
A SCUBA photometry observation of the other Ly$\,\alpha$ blob (`blob~2')
gives a $2.8\sigma$ detection of a much weaker sub-mm counterpart,
$S_{\rm 850}\,{=}\,3.3\,$mJy.
The 850\mum\ and 450\mum\ images of the `blob~1' are presented in Fig.~1,
convolved with a 14\arcsec\ Gaussian, 
while the fluxes at
various wavelengths for the two blob sources and surrounding sub-mm sources
are listed in Table~1. 
Other known $z\,{\sim}\,3$ objects from Steidel
et al.~(2000) are denoted with crosses on the 850\mum\ map,
while a \lya\ narrowband image with $K$-band contours and SCUBA error circles
overlaid is presented in Fig.~2 for both blobs.
The extremely luminous sub-mm source
appears coincident with the central Ly$\,\alpha$ knot (`blob~1').
Although this \lya\ peak remains undetected in the broadband optical to $R>26$,
it has a $K$-band counterpart  with an associated $R-K>5$. The sub-mm
centroid lies only 2 arcsec from this \lya\ peak, entirely consistent
with the JCMT pointing errors.
No radio counterpart was detected at the sensitivity limits of the VLA images
consistent with a high-$z$ emitting source.

For `blob~2', the SCUBA {\it photometry} observation was targeted on the
\lya\ center, near a weak $K$-band source,
where we identify a possible ${\sim}\,3\,$mJy detection.  However, without
higher resolution it is difficult to know which source within the
${\sim}\,15^{\prime\prime}$ beam is giving rise to the sub-mm emission.
In particular an LBG, M14, lies at the edge of the SCUBA beam, 
${\sim}\,$6\arcsec\ northeast of the `blob~2' center. If M14 was generating
the sub-mm flux, then the beam efficiency correction would imply an
850\mum\ flux of ${\sim}\,5\,$mJy, similar to the very red LBG ($R-K=4.6$)
detected by SCUBA (Westphal-MMD11 -- Chapman et al.~2000a).
However, M14 is a very faint 
LBG ($R=25.5$) and is not especially red ($R-K=3$), 
and is thus not expected to be a strong sub-mm emitter.  Moreover
the relatively faint sub-mm emission makes it unlikely that a radio counterpart
could be detectable for a $z\,{\sim}\,3$ in the archive VLA maps. 

\subsection{The other sub-mm sources}
As well as the central `blob~1' detection, there are perhaps as many as
three additional source detections ($\ga3\sigma\,{\sim}\,10$\,mJy)
in the 850\mum\ map.  Our best attempt to describe the extended
region to the west is as three separate sources in a blend (source 3
in the Table).  However, low signal-to-noise and resolution means
that this is little more than a guess.
The 850\mum\ blend region may be extended at 450\mum, but is detected only
in the northern part.
Two possible sources to the NW (4th and 5th in the table) appear unresolved,
but are near the edge where the noise is worse and poorly defined,
although source 6 appears coincident with a 450\mum\ source.

Radio observations provide rough constraints on the redshifts of
sub-mm sources (Carilli \& Yun 1999,2000; Barger et al.~2000). 
Our sub-mm detected galaxies in SSA22 have inferred bolometric luminosities
several times that of Arp~220,
with expected 1.4\,GHz radio fluxes ${\sim}\,85\,\mu$Jy (Helou et al.~1985)
and thus lie below the detection threshold of current VLA observations.
Applying the Carilli \& Yun (2000) relation to provide redshift
constraints from the 850\mum\ detections and $1.4\,$GHz upper limits shows
that all these sources have predicted redshifts $z\,{\ga}\,2.0$.

The ratio $S_{450}/S_{850}$ can also loosely constrain the source
redshifts (cf.\ Hughes et al.~1998). While the central blob source, and
source 4 are consistent
with $z\,{\sim}\,3$, the sources 3 and 5 in the Table with 450\mum\ 
detections have flux ratios suggesting lower redshifts.
For the western blend region, confusion makes it hard to draw any inferences.
The flux ratio for the northern source (850\mum\ source 5) suggests a
redshift less than 2. However, the source is near the edge,
identification of the 450\mum\ peak with
the 850\mum\ source is far from certain, and the
lack of an obvious radio counterpart suggests $z\,{>}\,2$.

Turning now to optical identifications, we find 
no known $z\,{\sim}\,3$ optical/near-IR
candidates within the error circles of these additional sub-mm
source detections. The sub-mm sources may be extremely faint ($R\,{>}\,26$,
$K\,{>}\,22$), 
similar to the radio selected sub-mm bright populations studied by 
Barger et al.~(2000) and Chapman et al.~(2000b).
However, not all sub-mm sources are associated with such optically
faint counterparts 
(e.g. the LBG W-MMD11, Chapman et al.~2000a). In addition, only
about half of the blue, UV bright $z\sim3.0$ galaxies are included
in LBG catalogs (largely due to photometric errors and blending
with foreground objects), thus it is possible
that a sub-mm source could be associated with a UV bright
$z\,{\sim}\,3.0$ galaxy even if it has no counterpart in our LBG list.
The relatively
low detection significance of these surrounding sub-mm sources, and the
absence of any obvious counterparts at other wavelengths, suggest that
we exercise caution in over-interpreting these objects.
Nevertheless, the number density of
${\sim}\,10\,$mJy SCUBA sources is about an order of magnitude higher
than in SCUBA blank fields (cf.\ Blain et al.~1999; Barger et al.~2000; 
Chapman et al.~2000c); at least some of these sources are likely
to be associated with the $z\,{=}\,3.09$ structure.

\section{Discussion and Conclusions}

The central `blob~1' SCUBA source is
among the most luminous high-$z$ sub-mm sources known.
It is especially interesting that this
object was discovered through the presence of tremendous amounts
of \lya\ emission, usually thought of as easily destroyed by the presence
of dust.  This begs the question: would other bright sub-mm
sources (if their redshifts were known) exhibit similar \lya\
`halos'?

One luminous sub-mm {\it selected} source, SMMJ02399-0136, is known to be
associated with an extended ($>$8\arcsec) \lya\ region (Ivison et al.~1998), 
although the actual extent has not yet been constrained through a
narrow-band image.
Both star forming and AGN components have been identified, although 
the primary power source for the sub-mm emission has not been constrained.
A few high-$z$ radio galaxies with giant \lya\ halos
have also been detected in the sub-mm, including 4C41.17
(Chambers et al.~1990) which has a possible `cluster' of sub-mm sources
associated (Ivison et al.~2000).
However, in the case of this SSA22 blob, the \lya\ halo is
hard to explain through excitation of the ambient medium by
a radio jet, since there is no known radio counterpart here.
Archibald et al.~(2000) have observed a sample of high
redshift radio galaxies using SCUBA, detecting several with known \lya\ halos.
The sub-mm emission from the `blob~1' object is more luminous than any
radio galaxy in this survey of 47 sources over
$1\,{<}\,z\,{<}\,4$.  Indeed, the lack of significant radio emission from
our blob~1 SCUBA object suggests that the \lya\ halos surrounding
distant radio galaxies may not be entirely related to the radio sources.

The questions of the ionizing source for the Ly$\,\alpha$ blobs and the
nature of the sub-mm emission remain.
The blobs are probably excited by either
photoionization from a central source,
or as a result of virialization in a deep potential. The sub-mm emission
could be either a UV-luminous star forming galaxy or an obscured AGN.
More extended sub-mm emission such as
from a cooling flow or even from the Sunyaev-Zel'dovich effect are
also possible.
The lack of a UV source in the optical images, coupled with the large
dust mass implied by the sub-mm detections, make 
direct photoionization of the \lya\ from such 
a central AGN or starburst difficult to imagine.
However, it is conceivable that complicated geometrical and radiative
transfer effects could conspire to allow substantial
\lya\ fluxes at the same time that the continuum sources are
heavily obscured by dust along our line of sight 
(e.g.~Neufeld 1991). 
The fact that the second \lya\ blob is not a strong sub-mm
source suggests that an obscured continuum source need not be
the primary driver for the \lya\ emission.

For a pure star-burst galaxy,
the star formation rate extrapolated
from the sub-mm would be in excess of $500\,h^{-2}{\rm M}_\odot{\rm yr}^{-1}$
for `blob~1'.  By analogy with the giant Ly$\,\alpha$ halos around high-$z$
radio galaxies, there is reason to suspect a luminous
AGN at the heart of the Ly$\,\alpha$ blob.  Although
the redshift of the \lya\ `blob~1' ($z\,{=}\,3.09$) is relatively high for
radio detection, the farIR/radio correlation for radio quiet sources
(Helou et al.~1995) suggests
a 1.4\,GHz flux density of ${\sim}\,170\,\mu$Jy if it is a steep spectrum
object (Carilli \& Yun 1999),
and possibly a comparable 8.5\,GHz flux if it had the
flatter spectrum characteristic of an AGN.
Our Bayesian 95\% upper limits of 142$\,\mu$Jy (1.4\,GHz) and 95$\,\mu$Jy
(8.5\,GHz) lie well within the uncertainties of the Carilli \& Yun (2000)
relation, and deeper
observations will be required to further constrain the properties of the system.
The ratio $S_{450}/S_{850}$ for the central source is $3.8\pm1.2$,
which is consistent with a starburst or radio quiet AGN lying at
$z\,{\simeq}\,3$ (cf.\ Hughes et al.~1998).

Another possibility is that the 850\mum\ flux could be partly
Sunyaev-Zel'dovich increment,  although the detection of 450\mum\ flux argues
against this being a significant component.  
On theoretical grounds, the surface brightness of the
increment for rich local clusters would be expected to be around
$1\,$mJy per SCUBA beam, and all other things being equal, S-Z surface
brightness is independent of redshift. 
The S-Z signal is approximately $\propto M\times\rho$:
the background density is proportional to $(1+z)^3$, 
and so the core of a cluster
virializing at high redshift would be much denser.
However, the characteristic non-linear mass in an
$\Omega_M=1$ universe goes like $(1+z)^{-6/(3+n)}$,
where $n$ is the power-spectrum slope.  For CDM
models $n$ is -1 to -2 on the relevant scales.
Hence, though the estimate is dependent on cosmology,
the simple $M\times\rho$ argument
suggests that the S-Z contribution is not likely to explain all our signal.
The expected {\it decrement\/} at $8.4\,$GHz would be
${\simeq}\,100\,\mu$Jy, and hence at the limit of being detectable currently.

If the blobs are associated with cooling-flow-like phenomena
(e.g.~Hu~1992), then some of the sub-mm emission could come from dust
associated with distributed star-formation in the cooling flow.
Edge et al.~(1999) and Chapman et al.~(2000c) have searched for such emission
in well-known rich clusters with null results.  However, local
cooling flow clusters are not necessarily a proper benchmark, given that the
associated gas densities could conceivably be much higher at $z\,{\sim}\,3$.
Fardal et al.~(2000) and Haiman et al.~(2000) have recently modelled the 
cooling radiation from forming galaxies and concluded that sub-virial 
temperatures for the gas could lead to much lower x-ray luminosities than 
one would expect from extrapolating local cooling flows.
While it is not
inconceivable that some of the flux has this source, even optimistic
estimates for the strength of the signal, as discussed in the above papers,
come out at a few mJy.  

Future observations will continue to elucidate the nature of the `blobs'.
For optically thin sub-mm emission, we can estimate the expected
x-ray signal of the `blob~1' sub-mm object
using different analogs for the central source at $z\,{=}\,3.09$,
and representative sub-mm/x-ray indices (cf.\ Fabian et al.~2000).
Even the most heavily obscured (Compton-thick) AGN, analogous to
NGC\,6240, should be detectable with Chandra or Newton
with soft and hard x-ray fluxes of order
${\simeq}10^{-18}{\rm W}\,{\rm m}^{-2}$, while
an Arp220 analog would be even fainter.
Deeper radio maps or higher frequency interferometer maps would be able
to distinguish any SZ component, as well as confirm the other sub-mm
sources and their positions.

The extreme sub-mm luminosity from the central \lya\ halo in `blob~1'
(and weaker emission detected from `blob~2')
is an intriguing phenomenon, given the fragility of the
\lya\ line in the presense of dust.  This combination of deep, large-area
\lya\ surveys and sub-mm images may be interesting to explore further as a
way of unearthing information about clusters in their early stages of
formation.

\acknowledgments
We thank the  staff of the JCMT
for their assistance with the SCUBA observations. 
The James Clerk Maxwell Telescope is operated by
The Joint Astronomy Centre on behalf of the Particle Physics and
Astronomy Research Council of the United Kingdom, the Netherlands
Organization for Scientific Research, and the National Research
Council of Canada.


%
%
\begin{figure}
\begin{center}
\epsfig{file=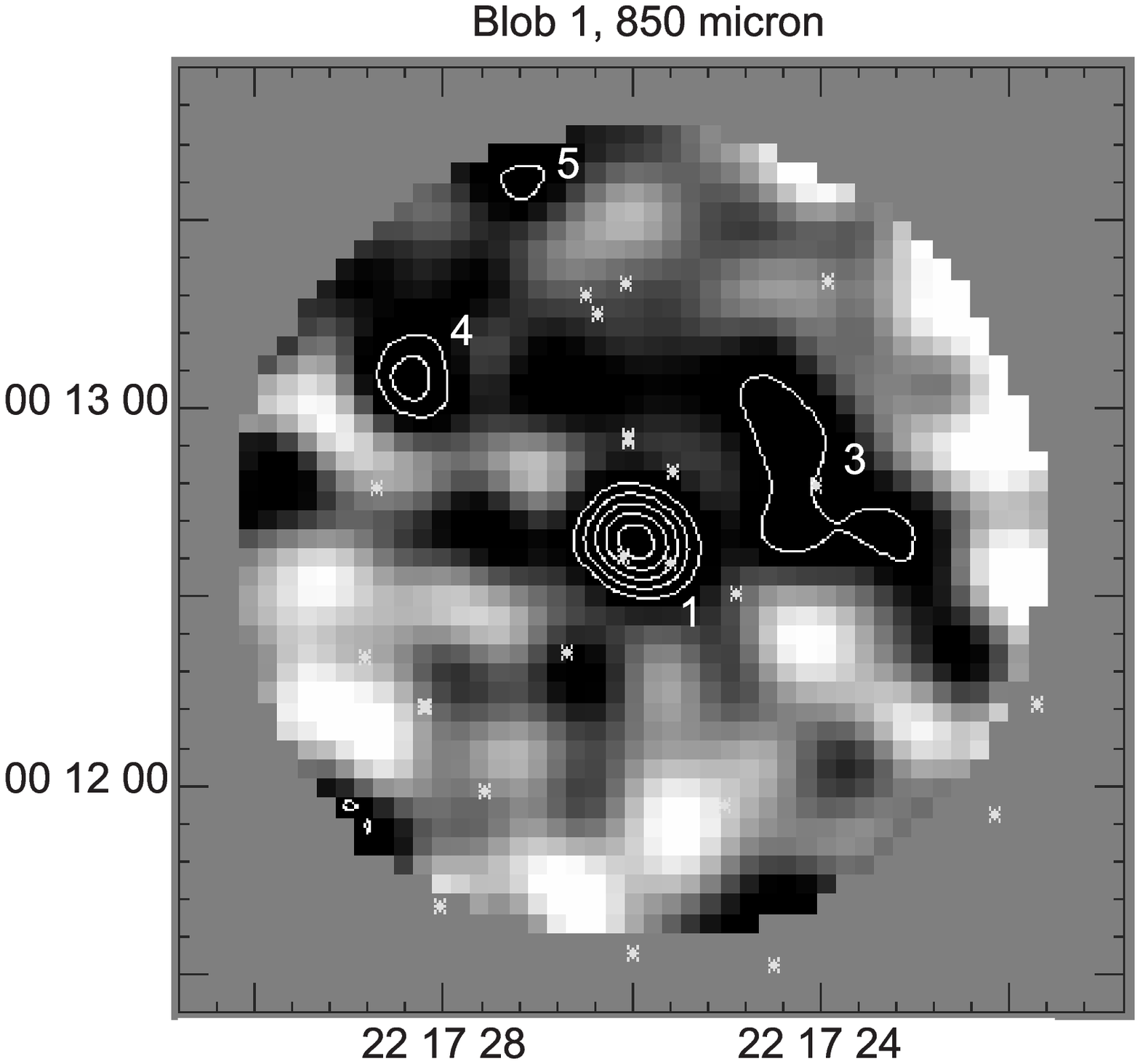,height=8.25cm,angle=0}
\epsfig{file=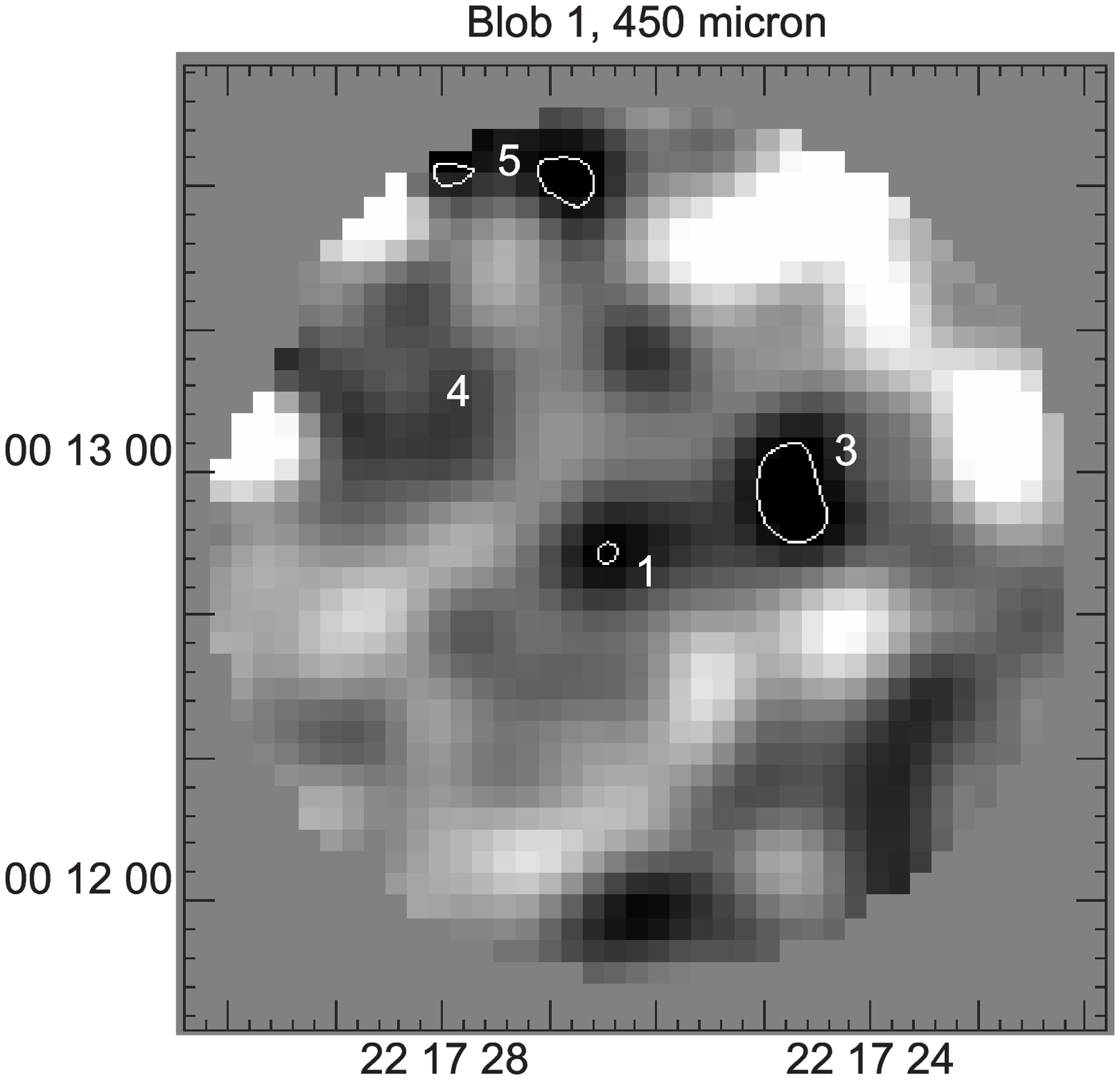,height=8.55cm,angle=0}
\end{center}
\figcaption[chapman.fig2]{The sub-mm 850\mum\ (left) and 450\mum\ (right) 
images of the extended \lya\ region.  These maps have been smoothed with
a 14\arcsec\ Gaussian beam and have contours at $3\sigma$, $4\sigma$,
$5\sigma$, \ldots. 
The noisy edges have been clipped, but possible detections very close
to the edge of the remaining images could still be noise artifacts.  Crosses on 
the 850\mum\ image denote known $z\sim3$ sources, discovered either
as LBGs or through the \lya\ narrowband imaging of Steidel et al.~(2000).
}
\label{f2}
\end{figure}

%
%
\begin{figure}
\begin{center}
\epsfig{file=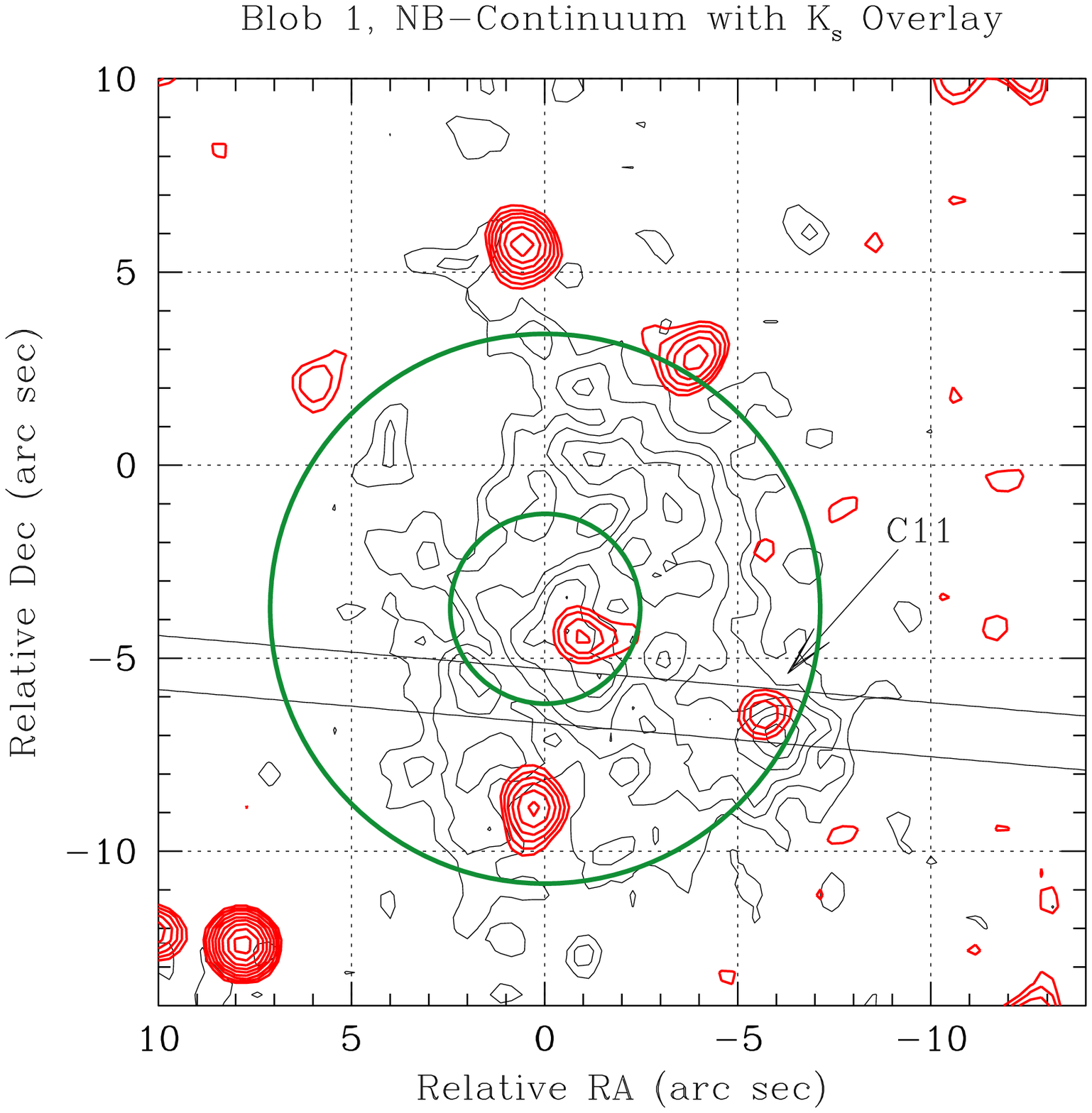,height=8.05cm,angle=0}
\epsfig{file=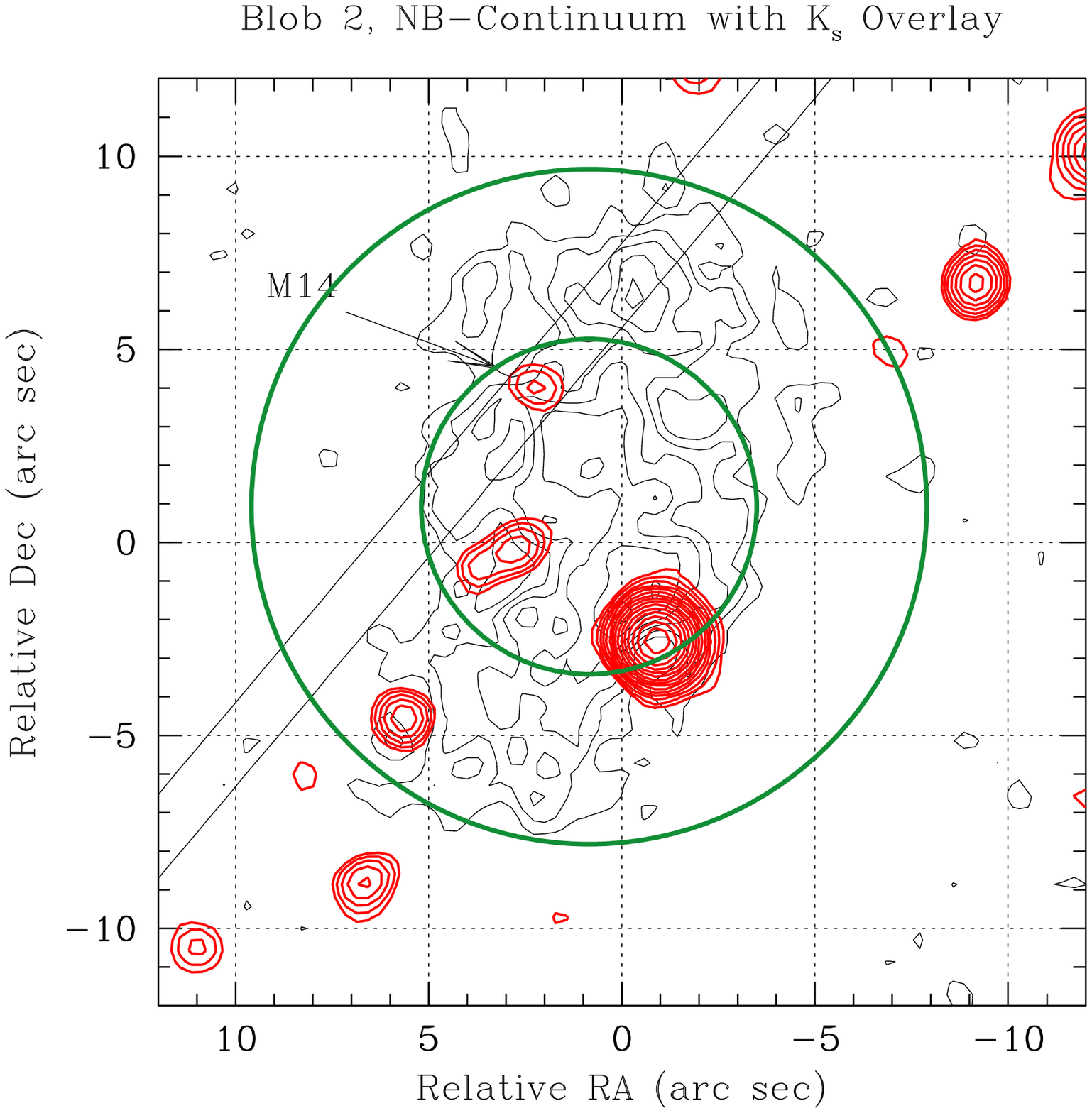,height=8.05cm,angle=0}
\end{center}
\figcaption[chapman.fig1]{
The lyalpha narrowband images of `blob~1'
(left) and `blob~2' (right) with NIRC $K-$band imagery (contours) superposed 
(as originally presented in Steidel et al.~2000). Known Lyman-break galaxies,
M14 and C11, are indicated.
Circles have been overlaid to indicate the expected sub-mm error circle
and beamsize (larger for the `photometry' observation of `blob~2' because
of the 2\arcsec\ dither pattern employed).}
\label{f1}
\end{figure}

%
%
\begin{deluxetable}{lrcccc}
\tablewidth{280pt}
\scriptsize
\tablenum{1}
\label{table-1}
\tablecaption{\sc \small Properties of sub-mm objects within the 2
Ly$\,\alpha$ `blob' regions\label{tab1}}
\tablehead{
\colhead{source} &  \colhead{$S_{850\mu{\rm m}}$}
 & \colhead{$S_{450\mu{\rm m}}$} & \colhead{$S_{1.4{\rm GHz}}$}
 & \colhead{$S_{8.5{\rm GHz}}$} 
 & \colhead{redshift$^{\rm b}$}\\
\colhead{} & \colhead{(mJy)} & \colhead{(mJy)} & \colhead{($\mu$Jy)} &
 \colhead{($\mu$Jy)}  }
\startdata
{\bf Blobs} & & & &\\
(1) SMMJ221726+0013& \pho20.1$\pm$3.3 & 76$\pm$24 & \pho35.4$\pm$59.2 & $-20.7\pm$48.3 & $>2.8$\\ 
(2) SMMJ221717+0015&  \pho3.3$\pm$1.2 & 29$\pm$15 & \pho55.4$\pm$59.7 &
 $-50.7\pm$51.1 & $>1.7$\\  
%
{\bf Other sources} & & & &\\
(3) SMMJ221724+0012&  \pho15.4$\pm$3.5$^{\rm c}$ & 95$\pm$29 & 100.7$\pm$61.7 &
 \Pho33.5$\pm$47.4 & $>2.4$\\ 
(4) SMMJ221728+0013& 12.2$\pm$3.7 & 41$\pm$33 & 214.2$\pm$47.8 & $-98.5\pm$45.7 & $>2.3$\\ 
(5) SMMJ221727+0014&  \pho10.7$\pm$3.8 & 87$\pm$32 & \pho44.2$\pm$44.9 &
 $-18.5\pm$43.7 & $>2.4$\\  
\enddata
\vspace*{-0.5cm}
\tablerefs{.\\
a) Sub-mm flux densities do not contain a ${\sim}\,$10\% calibration
uncertainty at 850\mum\ and ${\sim}\,$20\% at 450\mum.
The 450\mum\ measurements
are made in matched 15\arcsec\ apertures to the 850\mum\ centroids.\\
b) redshift lower limits estmated from the sub-mm/radio index using the
Bayesan 95\% radio upper limit and Carilli \& Yun (2000) indicator.\\
c) 20\arcsec\ diameter aperture estimate.
}
\end{deluxetable}

\end{document}